\documentclass{article}
\usepackage{amsmath}

\title{A short note on conflicting definitions of locality}
\author{Antonio Di Lorenzo
\\ Instituto de F\'{\i}sica, Universidade Federal de Uberl\^{a}ndia, \\ 38400-902 Uberl\^{a}ndia MG, Brasil}
\begin{document}
\maketitle

\begin{abstract}
There are various non-equivalent definitions of locality. 
Three of them, impossibility of instantaneous communication, impossibility of action-at-a-distance, and impossibility of faster-than-light travel, while 
not fully implying each other, have a large overlap. When the term non-locality is used, most physicists think that one of these three conditions is being violated. There is a minority of physicists, however, who uses ``locality" with a fourth meaning, 
the satisfaction of the hypotheses underlying Bell inequality. 
This definition devoids Bell's theorem of its profoundness, reducing it to a mere tautology. 
It is demonstrated here, through a classical example using a deck of cards, that this latter definition of ``locality" is untenable. 
\end{abstract}

Bell inequality \cite{Bell1964}, and the related inequalities derived later \cite{Clauser1969,Clauser1974}, 
constitute an important result in theoretical physics, since their experimental violation \cite{Aspect1982a,Aspect1982b} 
binds the shape of any possible physical theory. At present, quantum mechanics explains 
satisfactorily the experimental data, but, even if in the future it should be substituted 
by a new theory, the observed violation of Bell inequality (BI) requires such hypothetical new theory 
to satisfy some general properties. 
It is commonly encountered the statement that the violation of BI implies that reality is non-local. 
As far as I know, locality can have any of the following meanings
\begin{description}
\item[a.] Impossibility of instantaneous communication between two space-like separated parties (no-signaling).
\item[b.] Impossibility of instantaneous action-at-a-distance. 
\item[c.] Impossibility for a body to have an arbitrary velocity.
\end{description}
These three formulation are non-equivalent, and one should specify which one is intended when using the term ``locality". 
For instance, the violation of (c) certainly implies the violation of (a) and (b), but the \emph{vice versa} is not true. 
I believe that the majority of physicists has in mind one of these three possibilities when hearing ``locality", as practically the totality of non-physicists. 
However, a fourth meaning has been attached to ``locality" by a minority of physicists. This leads to misunderstandings  
and to discussions leading nowhere among physicists, and also to a sense of bewilderment of the wide public towards quantum mechanics, 
which is already a counter-intuitive theory by itself and whose exposition need not be further obscured by misleading terminology.  
This fourth definition of non-locality refers specifically to a theory predicting events $e_j$ observed by local detectors $\Sigma_j$ at space-like separated regions $R_j$ given that a system is specified by some parameters $\lambda$, 
\begin{description}
\setcounter{enumi}{3}
\item[d.] A theory is local if the probability of observing the events $e_j$ factorizes as 
\begin{equation}\label{eq:fact}
P(\{e\}|\lambda,\Sigma)=\prod_j P(e_j|\lambda,\Sigma_j)
\end{equation} 
\end{description}
Following Ref.~\cite{Fine1982a}, we shall refer to the property described in Eq.~\eqref{eq:fact} as factorability (or factorizability). 
In the special case of a bipartite entangled two-level system, we have that $\Sigma_j$ represent (pseudo)spin components, and $e_j=\pm 1$ in appropriate 
units. The hypothesis of factorability coincides then with one of the hypotheses at the basis of Bell inequality (the other hypothesis being that $\lambda$ 
are distributed independently of the setup $\Sigma$).
We want to show that if one takes factorability as a definition of  locality then various models which are patently local according to any 
of the definitions (a)-(c) are classified as non-local according to (d). 
Let us first rewrite the left hand side of Eq.~\eqref{eq:fact} by applying repeatedly Bayes's theorem
\begin{equation}\label{eq:factbayes}
P(\{e\}|\lambda,\Sigma)=\prod_j P(e_j|E_j,\lambda,\Sigma),
\end{equation} 
where $E_1=\emptyset$ is the empty set, and $E_j=\{e_1,e_2,\dots,e_{j-1}\}$. 
Eq.~\eqref{eq:factbayes} is a mathematical identity following from the rules of probability theory, and does not involve any physical assumption. 
Now, let us consider the first factor in the right hand side of Eq.~\eqref{eq:factbayes}, $P(e_1|\lambda,\Sigma)$. 
If we invoke locality either in the form (a) or (b), we have the physical equality $P(e_1|\lambda,\Sigma)=P(e_1|\lambda,\Sigma_1)$. 
The second factor, however, is 
$P(e_2|e_1,\lambda,\Sigma)$. Invoking again (a) or (b), we have that $P(e_2|e_1,\lambda,\Sigma)=P(e_2|e_1,\lambda,\Sigma_1,\Sigma_2)$. 
Analogously, we have in general that 
\begin{equation}
P(e_j|E_j,\lambda,\Sigma)=P(e_j|E_j,\lambda,\Sigma_1,\dots,\Sigma_j).
\end{equation} 
Thus, by applying (a) or (b) we cannot justify Eq.~\eqref{eq:fact}. In other words, condition (d) is stronger than (a) or (b). 
This was pointed out by Jarrett \cite{Jarrett1984}. In order to obtain the equality in Eq.~\eqref{eq:fact} 
we have to make a further hypothesis, which Jarrett referred to as completeness \cite{Jarrett1984}, 
and Shimony as outcome-independence \cite{Shimony1990}. We shall use the latter terminology, since it is purely technical and thus 
immune to misinterpretations. This hypothesis is simply that
\begin{equation}\label{eq:oi}
P(e_j|E_j,\lambda,\Sigma_1,\dots,\Sigma_j)=P(e_j|\lambda,\Sigma_1,\dots,\Sigma_j) ,
\end{equation}
i.e., the conditional probability of observing $e_j$ given that $e_1,\dots,e_{j-1}$ were observed is identical 
to the marginal probability of observing $e_j$. 
Now, while in order to establish $P(e_j|E_j,\lambda,\Sigma_1,\dots,\Sigma_j)$ it is necessary that observers 
in $R_1,\dots,R_{j-1}$ communicate their results to the observer in $R_j$, the marginal $P(e_j|\lambda,\Sigma_1,\dots,\Sigma_j)$ 
can be determined by means of a local measurement in $R_j$, without need for communication. 
Then, it is now possible to invoke once again locality in either form (a) or (b) and obtain finally 
\begin{equation}\label{eq:oiplusloc}
P(e_j|E_j,\lambda,\Sigma_1,\dots,\Sigma_j)= P(e_j|\lambda,\Sigma_j),
\end{equation}
which, upon substitution in the right hand side of Eq.~\eqref{eq:factbayes} yields Eq.~\eqref{eq:fact}. 
So far, we have basically reformulated the conclusions of Ref.~\cite{Jarrett1984}, that, however, is ignored by most physicists, so we hope that 
this short note help propagating this important result. 

Let us establish sufficient conditions for the validity of outcome-independence as formulated in Eq.~\eqref{eq:oi}. 
Determinism is a sufficient condition: if the knowledge of $\lambda$ determines the outcome of the measurements $\Sigma_j$, all other 
information is redundant.\footnote{Historical note: 
Determinism also implies that $P(e_j|\lambda,\Sigma_j)=0$ or $1$. 
Refs.~\cite{Bell1964,Clauser1969} actually assumed locality, in the form (a) or (b), and determinism. The fact that 
$P(e_j|\lambda,\Sigma_j)\in\{0,1\}$ was exploited in both papers. 
Later on \cite{Clauser1974}, it was realized that the inequalities could still be derived if $0\le P(e_j|\lambda,\Sigma_1,\dots,\Sigma_j)\le 1$, 
thus it was claimed that the hypothesis of determinism was dropped, and that Bell inequality established the incompatibility of all local stochastic theories with quantum mechanics. However, in deriving all Bell-type inequalities, a consequence of determinism, namely the hypothesis of outcome-independence, 
was retained implicitly, and this was realized only ten years later \cite{Jarrett1984}. Meanwhile, so strong was the belief that the hypotheses underlying Bell inequality only required locality, that the requirement (d) was taken as the definition of locality.}
Another sufficient hypothesis is that of probabilistic determinism, which is not an oxymoron, but 
means that knowledge of the $\lambda$ determines the probability of the outcome, not the outcome itself. 
Perhaps this concept coincides with what Jarrett \cite{Jarrett1984} calls ``completeness'', which we avoid since it is a term laden 
with subjective meanings. 
Another sufficient condition is separability \cite{Howard1989}: the parameters $\lambda=\bigcup_j \lambda_j$, where $\lambda_j$ 
are parameters attached to the particle number $j$ and represent prepossessed values. Then we have that the probability of any outcome $e_j$
can be determined by knowledge of the $\lambda_j$ alone, i.e., 
$P(e_j|E_j,\lambda,\Sigma_1,\dots,\Sigma_j)=P(e_j|\lambda_j,\Sigma_j)$, which is a special form of outcome-independence. 
Finally, it has been proved recently \cite{Hall2011} that if a model satisfies factorability then there is a natural extension of this model which is 
deterministic. For convenience, the proof is reported in the appendix. 

In order to show the absurdity of identifiying Eq.~\eqref{eq:fact} with locality, we shall provide now a simple model which is manifestly local, but  
would be classified as non-local according to (d). 
Let us consider the following experiment: 
a dealer prepares two decks of cards which can be a King ($K$) or a 
Queen ($Q$), and Black ($B$) or Red ($R$); each deck is 
subdivided into pairs, such that each pair is formed by a King and 
a Queen, and a Black and a Red card. 
In the first deck ($D_1$), 30\% of the pairs are $(KR,QB)$, and 70\% $(KB,QR)$. 
In the second deck ($D_2$), 70\% of the pairs are $(KR,QB)$, and 30\% $(KB,QR)$. 
The dealer chooses a deck at random, with equal probability, then extracts a pair out of the deck, 
and handles one card each to two observers, one, $L$, sitting to his right and the other, $R$, to his left. 
In this model, the hidden variable $\lambda$ is the deck which has been chosen. 
Once the pair is extracted from the deck, $\lambda$ is not localized on either card, but is a global property, which cannot 
be reconstructed by observers $L$ and $R$ by making local observations (even if they compare their results). 
The only way to determine $\lambda$ would be to check which deck was chosen at the location of the preparation. 
Now let us consider the 
joint probability that $L$ will receive a King and $R$ will receive a black card
given that deck 1 was chosen: 
\begin{equation}
P(K,B|\lambda=D_1)= \frac{3}{20} ,
\end{equation}
since a pair with a red King and a black Queen will be extracted with probability $3/10$, and the black Queen is received by $R$ half of the times.  
On the other hand, if factorability holds, we should have 
\begin{equation}
P(K,B|D_1)= P(K|D_1)P(B|D_1)=\frac{1}{4}. 
\end{equation}
Thus, according to definition (d), the model we have presented is non-local!  
The reader can surely make up uncountable other examples in which knowledge of some additional parameters does not break the correlations, 
theories which (d) classifies as non-local but are local according to one of (a)-(c). 

Notice that, with our example (or with any example which we can make out with our 
classical imagination), it is not possible to violate 
Bell inequality. The reason lies in the fact that in any classical 
case it is possible to find a more complete set of parameters 
$\lambda_{det}$ which determine univocally the outcome of each 
observation. 
E.g., in our case $\lambda_{det}$ would be determined by the knowledge 
of the pair of cards, and of which card goes to which observer. 
The peculiarity of quantum mechanics consists in the fact that no 
such complete characterization of a system exists in principle. 

In conclusion, we have proved that the definition (d) of locality conflicts with any other accepted definition of the term, and thus is untenable. 
I hope that this statement is not met with ideological animosity, 
but that the physics community may achieve unanimity over the definition of such an important 
concept. 
\section*{Appendix}
We reproduce the proof of the theorem in Ref.~\cite{Hall2011} according to which any factorable model admits 
a natural more fundamental parametrization which is deterministic. 
We recall that we are considering the detection of $N$ events at as many space-like separated regions. 
We take the events to be discrete, so that at region $R_j$ the possible values of the outcomes are 
$e_j\in\{o^{(j)}_1,o^{(j)}_2,\dots,o^{(j)}_{m_j}\}$. 
By hypothesis, the probability satisfies the factorability condition Eq.~\eqref{eq:fact}.
In addition to the parameters $\lambda$ we introduce the parameters $\mu$,  an $N$-uple of numbers uniformly 
distributed within the unit hypercube $[0,1]^{N}$. The event $e_j$ consists with certainty in the outcome 
$o^{(j)}_{k}$ when $\mu_j\in[\sum_{k'=1}^{k-1} P(o^{(j)}_{k}|\lambda,\Sigma_j),\sum_{k'=1}^{k} P(o^{(j)}_{k}|\lambda,\Sigma_j)]$, i.e., 
\begin{equation}
P(\{o^{(j)}_{k_j}\}|\lambda,\mu,\Sigma)=
\begin{cases}
1,&\!\!\!\!\!\text{if } \mu_j\!\in\!\left[\sum_{k'=1}^{k-1} P(o^{(j)}_{k}|\lambda,\Sigma_j),\sum_{k'=1}^{k} P(o^{(j)}_{k}|\lambda,\Sigma_j)\right],\\
0,&\!\!\!\!\! \text{otherwise}.
\end{cases}
\end{equation}
Upon integrating over $\mu$, Eq.~\eqref{eq:fact} is recovered.

\end{document}